\documentclass[onecolumn,11pt,draftcls]{IEEEtran} 
\usepackage{graphicx} 
\usepackage{bm} 
\usepackage{url} 
\usepackage{amsfonts,amssymb,amsmath}  
\usepackage{mystyle}

\allowdisplaybreaks
\begin{document} 
\title{Bilayer Protograph Codes for Half-Duplex Relay Channels} 
 
\author{Thuy Van Nguyen, {\em Student Member, IEEE}, Aria Nosratinia,
  {\em Fellow, IEEE}, and Dariush Divsalar, {\em Fellow,
    IEEE}
  \thanks{This work was presented in part in the 2010 IEEE
    International Symposium on Information Theory.}\thanks{
    Thuy Van Nguyen and Aria Nosratinia are with the Department of
    Electrical Engineering, The University of Texas at Dallas,
    Richardson, TX 75083-0688 USA, Email nvanthuy@utdallas.edu,
    aria@utdallas.edu. Dariush Divsalar is with the Jet Propulsion
    Laboratory, California Institute of Technology, Pasadena, CA
    91109-8099 USA Email: Dariush.Divsalar@jpl.nasa.gov.}}
 
\maketitle 
 
\begin{abstract} 

Despite encouraging advances in the design of relay codes, several
important challenges remain. Many of the existing LDPC relay codes are
tightly optimized for fixed channel conditions and not easily adapted
without extensive re-optimization of the code. Some have high encoding
complexity and some need long block lengths to approach capacity. This
paper presents a high-performance protograph-based LDPC coding scheme
for the half-duplex relay channel that addresses simultaneously
several important issues: structured coding that permits easy design,
low encoding complexity, embedded structure for convenient adaptation
to various channel conditions, and performance close to capacity with
a reasonable block length.  The application of the coding structure to
multi-relay networks is demonstrated. Finally, a simple new
methodology for evaluating the end-to-end error performance of relay coding
systems is developed and used to highlight the performance of the
proposed codes.
\end{abstract} 
 

\section{Introduction}  
 
To harvest the cooperative gain promised by information
theory~\cite{Cover1979,Kramer-Gupta:IT05}, coding schemes are needed
that can approach the fundamental limits of the relay channel.  Among the
early coding results for the relay channel were distributed coded
diversity schemes~\cite{Aria-Janani:TSP04,
  Aria-Hunter:ISIT02,Valenti-Zhao:VTC03}, which proposed convolutional
and turbo codes for the fading channel under cooperation.  Later on,
Duman et al.~\cite{Zheng-Duman:ISIT05,Zheng-Duman:Tcom05} proposed
turbo codes for the decode-forward (DF) relay channel in half-duplex
and full-duplex modes.  There is a large body of work focused on
designing LDPC codes for the relay
channel~\cite{Azmi:ISIT2008,Azmi:ISIT2009,Hu2006,Razaghi2007,Li2008,Chakrabarti2007,Cances2009}. These
works mostly utilize irregular LDPC codes and use density evolution
(or related) techniques to search for optimized irregular LDPC
ensembles operating at two different rates. Razaghi and
Yu~\cite{Razaghi2007} produce bilayer LDPC code structures for the DF
scheme. This approach has been refined in other works
including~\cite{Azmi:ISIT2008,Azmi:ISIT2009}. Other examples of LDPC
codes for the half-duplex relay channel
include~\cite{Hu2006,Li2008,Chakrabarti2007,Cances2009}.

\begin{figure*} 
\centering 
\includegraphics{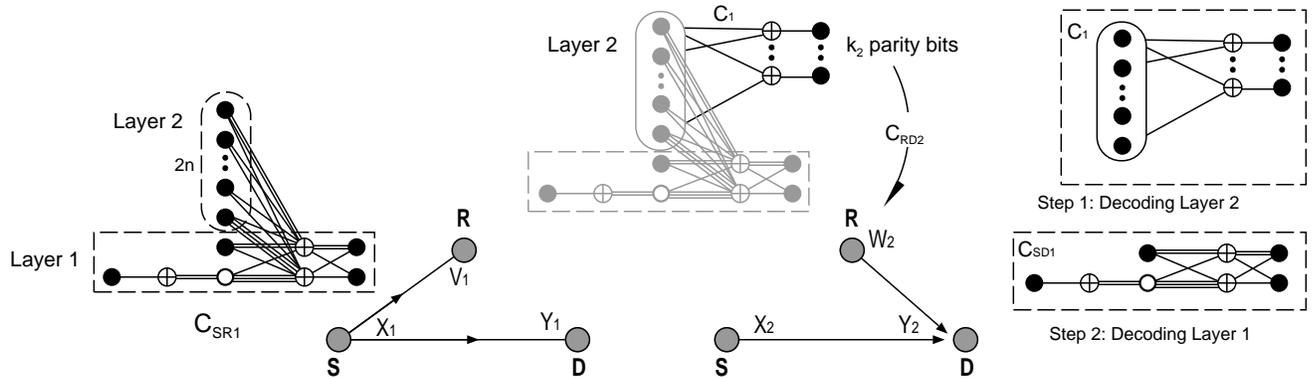} 
\caption{Bilayer-lengthened protograph design for half-duplex relay with 
 $R_{SD_1}=\frac{1}{2}$ and $R_{SR_1}=\frac{n+1}{n+2}$, $n=1,2,\ldots$} 
\label{fig:relay_BL} 
\end{figure*}

It has been known that the basic problem of coding for the DF relay
channel in the power-limited regime can be reduced to the following: a
code and its sub-code must be designed simultaneously that operate
with two different rates at two different SNRs at the relay and the
destination, respectively. Despite some progress, LDPC codes for the
relay channel are often painstakingly optimized to match to a set of
channel conditions without a structure to facilitate the optimization
of design, and many of them do not offer easy encoding.  This paper
proposes a class of LDPC relay codes that address three important
issues in an integrative manner: low encoding complexity, modular
structure allowing for easy design, and rate-compatibility so that the
code can be easily matched to a variety of channel conditions without
extensive re-optimization. In addition, the proposed codes offer
excellent performance.

In this paper, we concentrate on the DF protocol.  For
practical purposes, we further limit ourselves to the half-duplex
relaying where the relay cannot transmit and receive at the same time.
The main contribution of this paper is a coding scheme for the relay
channel using protographs~\cite{Thorpe2003} that are built with
bilayer expurgated and bilayer-lengthened
structures~\cite{Razaghi2007}. A protograph code~\cite{Thorpe2003} is
an LDPC code that can be constructed by extension from a small
bipartite graph called a protograph, whose graph topology is
represented by a so-called proto-matrix. Protograph codes can achieve
very good thresholds with low encoder complexity as well as fast
decoding~\cite{Thorpe2004,Divsalar2009}.  The proposed approach offers
flexibility in designing a family of rate-compatible embedded codes
for relay channels. These embedded codes allow a coding scheme whose
rate can easily be adapted to channel conditions, and are thus
suitable for designing multi-relay coding systems. We also introduce a
methodology to evaluate the end-to-end error performance of relay coding
schemes, and demonstrate the end-to-end performance of our proposed codes.



\section{Background} 
\label{sec:background}
\subsection{System Model} 
\label{sec:system} 

A half-duplex single-relay channel is shown in
Figure~\ref{fig:relay_BL}. 
$X_i$ and $W_i$
denote the transmitted signals from the source (S) and the relay (R)
while $Y_i$ and $V_i$ denote the received signals at the destination
(D) and relay, respectively. Subscript $i=1$ denotes the broadcast mode
which is active for a fraction $t$ of the transmission interval, and
subscript $i=2$ denotes the multiple access mode which is active for a
fraction $t'=1-t$ of the transmission interval. The received signals
are:
\begin{align} 
V_1 &= h_{SR}X_1 + N_{R_1} \nonumber\\ 
Y_1 &= h_{SD}X_1 + N_{D_1} \nonumber\\ 
Y_2 &= h_{SD}X_2 + h_{RD}W_2 + N_{D_2} \; ,
\label{eq:model} 
\end{align} 
where $h_{SD}$ and $h_{RD}$ are the S-to-D and R-to-D channel
coefficients, respectively, $N_{R_1}$, $N_{D_1}$ and $N_{D_2}$ are the
noise samples at the relay and the destination in the first and
second time slots, respectively.  All noise samples are assumed to
be Gaussian with zero mean and unit variance.  $P_{S_1} = E[X_1^2]$
represents the source transmission power in the first time slot
(duration $t$). Similarly $P_{S_2}$ and $P_{R_2}$ represent the source
and relay transmission powers in the relayed time slot (duration
$t'=1-t$). We also define $SNR_{SR}$, $SNR_{SD}$ and $SNR_{RD}$ as
signal-to-noise ratios by the relay and destination during the first
time slot, and by the destination during the second time slot,
respectively.

\subsection{Coding for Half-Duplex Relay Channels}

In the first time slot, a source sends a high-rate $C_{SR_1}$ code to
a relay and a destination, using a rate that is decodable at the
relay. In the second time slot, the relay transmits additional mutual
information to help the destination decode, producing an effectively
lower-rate $C_{SD_1}$.\footnote{\label{footnote:2}In our formulation
the source may also be active in the second time slot, but it will be
constrained to synchronously transmit the same signal/codebook as the
relay. Thus in this work the distinction between orthogonal and
non-orthogonal transmission is only in the received SNR during relay
time slots. Since we parameterize our analysis by received SNR without
any direct reference to source and relay powers, both orthogonal and
(synchronous) non-orthogonal transmissions are subsumed in the
following developments.} Thus the relay coding problem consists of a
simultaneous design of two codes that operate at two different SNRs
$SNR_{SR}>SNR_{SD}$, such that one is a subset of the other. This
problem has been attempted a number of times by forceful optimization,
but a more streamlined approach is now available via bilayer
expurgated and bilayer-lengthened LDPC
structures~\cite{Razaghi2007}. In the bilayer LDPC structure, either
the bit nodes or the check nodes are divided into two sets (called
layers) and the graph, although still a bipartite graph, is re-drawn
with two rows of check nodes sandwiching the bit nodes, or vice
versa~\cite[Fig. 3 and 6]{Razaghi2007}. This provides a convenient way
to build and illustrate subcodes, which we use for our purposes. For a
more comprehensive treatment and justification of the bilayer
structure please see~\cite{Razaghi2007,Chakrabarti2007}.

\subsection{Protograph Codes} 
\label{sec:protograph}

\begin{figure} 
\centering \includegraphics[width=2.55in]{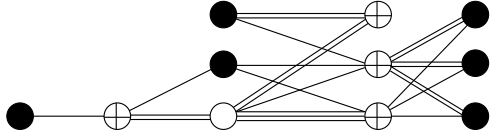} 
\caption{The rate-$1/2$ protograph with a threshold of 0.439 dB} 
\label{fig:code_439} 
\end{figure}

A protograph~\cite{Thorpe2003} is a Tanner graph with a relatively small
number of nodes. A protograph code is an LDPC code constructed from a
protograph by a copy-and-permute operation, where the protograph is
copied $N$ times and the $N$ variable-to-check pairs (edges)
corresponding to the same edge type of the original protograph are
permuted. Thus the protograph code has the same rate and the same node
distribution as the original protograph. Several capacity-approaching
protograph code designs have been proposed
in~\cite{Divsalar2009,El-Khamy2009}. Figure~\ref{fig:code_439}
shows a rate-$1/2$ protograph from~\cite{Thuy:ISIT10} consisting of 7
variable nodes and 4 check nodes that are interconnected by 24 edge
types. The dark circles represent transmitted variable nodes, the white
circle is a punctured node and the circles with a plus sign are parity
check nodes. This protograph can be represented by the proto-matrix
\begin{equation} 
H_{1/2} = 
\begin{pmatrix}
1 & 2&0 & 0 & 0 & 1 & 0 \\
0 & 3&1 & 1 & 1 & 1 & 0 \\ 
0 & 1&2 & 2 & 2 & 1 & 1 \\ 
0 & 2&0 & 0 & 0 & 0 & 2  
\end{pmatrix}
\; ,
\label{eq:B12_new} 
\end{equation} 
where the rows and columns represent the check nodes and variable nodes
respectively. The entries of the matrix represent the number of parallel
edges that connect the variable node and the check node. Using the PEXIT
technique~\cite{Liva2007}, the threshold of this code is $0.439$ dB, a
gap of $0.252$ dB of capacity. The linear minimum distance of this code
grows linearly with codelength\footnote{If a random permutation per each edge of
its protograph is used.} which is necessary for avoiding an error
floor~\cite{Divsalar_ITA10}.  

\section{Design of bilayer protograph-based LDPC codes for the relay channel} 
\label{sec:bilayer} 

\subsection{Bilayer Lengthened Structure}
\label{sec:bilayerL} 
A bilayer lengthened structure for the relay channel~\cite{Razaghi2007}
is shown in Fig.~\ref{fig:relay_BL} where a high-rate code is
constructed by adding variable nodes to the graph of a low-rate
code. Geometrically, the overall graph contains two layers (sets) of
variable nodes and one set of check nodes. One layer of variables
connecting to the checks forms a graph of a lower-rate code, and the
entire graph constitutes a high-rate code.  The corresponding parity
check matrix has the form:
\begin{equation} 
H_{SR_1}=[H_{SD_1} H_{e}] \; ,
\label{eq:BL_matrix} 
\end{equation} 
where $H_{SD_1}$ is the parity matrix of the lower-rate code which is
equivalent to the sub-code (Layer 1 or $C_{SD_1}$) and $H_e$ is the
extension matrix representing Layer 2. Layers 1 and 2 combined create
a capacity-approaching code for the source-relay link ($C_{SR_1}$) as
shown in Fig.~\ref{fig:relay_BL}. The relay decodes this codeword,
then protects the codeword bits in Layer 2 via $k_2$ parity bits.
These parities are encoded with $C_{RD_2}$ and transmitted to the
destination during the second time slot. The destination, using these
$k_2$ parity bits, can reliably detect the bits in Layer 2, therefore
they can be eliminated from the code graph. The remaining graph
contains just Layer 1, which constitutes a capacity approaching code
at the SNR of the SD link, which is decodable at the destination.

We will describe in the following an example for the design of the
bilayer lengthened codes with rates $R=\frac{n+1}{n+2}$,
$n=1,2.\ldots$, as an extension of the rate-$1/2$ protograph
of Eq.~\eqref{eq:B12_new}. For each new code, three variable nodes (or
columns) are added, as follows:
\begin{equation}
H_{\frac{n+1}{n+2}}= \left(H_{\frac{n}{n+1}}
\begin{array}{|ccc}
y_1 & y_2 & y_3 \\\hline
x_1 & x_4 & x_7\\
x_2& x_5 & x_8\\
x_3 & x_6 & x_9
\end{array}
\right) \; ,
\label{eq:B23_x}
\end{equation}
where $x_i$ and $y_i$ variables denote the number of parallel edges in
the extension graph, to be determined. Variables $y_i$ in the first
row correspond to the check node that connects to the degree-$1$
variable node. Variables $x_i$ in rows $2$-$4$ correspond to the
remaining check nodes. In order to preserve the linear growth of
minimum distance for the bilayer codes, the column sums in rows
$2$-$4$, namely sums of columns designated with variables $x_i$,
should be $3$ or higher~\cite{Divsalar_ITA10}.

We further simplify the problem by setting the maximum number of
parallel edges to $2$ (i.e. $x_i,y_i=0,1,2$). These constraints limit
the range of parameters thus simplifying the optimization. Our cost
function is the threshold which is calculated via the PEXIT
method. For this example, the best threshold is given by:

\begin{align*} 
H_{2/3}= \left(H_{1/2} 
\begin{array}{|ccc} 
0 & 1 & 1 \\ 
1 & 1 & 1\\ 
2& 1 & 2\\ 
0 & 1 & 0 
\end{array} 
\right)
\quad
H_{3/4}= \left(H_{2/3} 
\begin{array}{|ccc} 
0 & 0 & 2 \\ 
2 & 2 & 0\\ 
1  & 1 & 2\\ 
0 & 0 &1 
\end{array} 
\right)
\\ 
H_{4/5}= \left(H_{3/4} 
\begin{array}{|ccc} 
0 & 1& 2 \\ 
1 & 2& 2\\ 
2 & 1& 1\\ 
0 & 0& 0 
\end{array} 
\right)
\quad
H_{5/6}= \left(H_{4/5} 
\begin{array}{|ccc} 
0 & 0 &1 \\ 
2 & 2 &0\\ 
1 & 1 &2\\ 
0 & 0 &2 
\end{array} 
\right)
\\ 
H_{6/7}= \left(H_{5/6} 
\begin{array}{|ccc} 
0 & 0 &1 \\ 
1 & 2 &1\\ 
2 & 1 &2\\ 
0 & 1 &0 
\end{array} 
\right)
\quad
H_{7/8}= \left(H_{6/5} 
\begin{array}{|ccc} 
0 & 0&2 \\ 
2 & 2&0\\ 
1 & 1&2\\ 
0 & 0&2 
\end{array} 
\right)
\\ 
H_{8/9}= \left(H_{7/8} 
\begin{array}{|ccc} 
0 & 0&0 \\ 
0 & 1&2\\ 
2 & 2&1\\ 
1 & 1&0 
\end{array} 
\right)
\quad
H_{9/10}= \left(H_{8/9} 
\begin{array}{|ccc} 
0 & 0&2 \\ 
1 & 2&0\\ 
2 & 1&2\\ 
0 & 0&2 
\end{array} 
\right)
\end{align*} 

The iterative decoding thresholds of these codes calculated by
PEXIT technique~\cite{Liva2007} are given in Table~\ref{ta:proto}. For rates $>2/3$,
the produced codes have iterative decoding thresholds within $0.1$ dB of
the capacity, and the rate-2/3 code has a threshold within $0.152$ dB.
\begin{table}
\caption{Thresholds of proposed protograph codes} 
\centering 
\begin{tabular}{|c|c|c|c|} 
\hline 
Code & Protograph& Capacity & Gap to\\ 
Rate & threshold (dB) & threshold (dB) & capacity\\ 
\hline 
\multicolumn{4}{|c|}{Bilayer lengthened design}\\ \hline
1/2 & 0.439 & 0.187 & 0.252\\  
2/3 & 1.223 & 1.059 & 0.164\\  
3/4 & 1.720 & 1.626 & 0.094\\  
4/5 & 2.136 & 2.040 & 0.096\\  
5/6 & 2.455 & 2.362 & 0.093\\  
6/7 & 2.718 & 2.625 & 0.093\\   
7/8 & 2.941 & 2.845 &0.099 \\   
8/9 & 3.125 & 3.042 &0.083 \\   
9/10 & 3.295 & 3.199&0.096 \\ \hline 
\multicolumn{4}{|c|}{Bilayer expurgated design}\\
\hline
3/4   & 1.720 & 1.626 & 0.094\\  
2/3   & 1.182 & 1.059 & 0.123\\  
7/12 & 0.809 & 0.590 & 0.219 \\  
1/2   & 0.420 & 0.187 & 0.233\\   
5/12 & 0.144 & -0.185 &0.329 \\  
1/3 & -0.263 & -0.497 &0.234 \\ \hline 
\end{tabular} 
\label{ta:proto} 
\end{table}

\subsection{Design of Bilayer Expurgated Protograph Codes} 
\label{sec:bilayerEx} 

\begin{figure*}
\centering 
\includegraphics{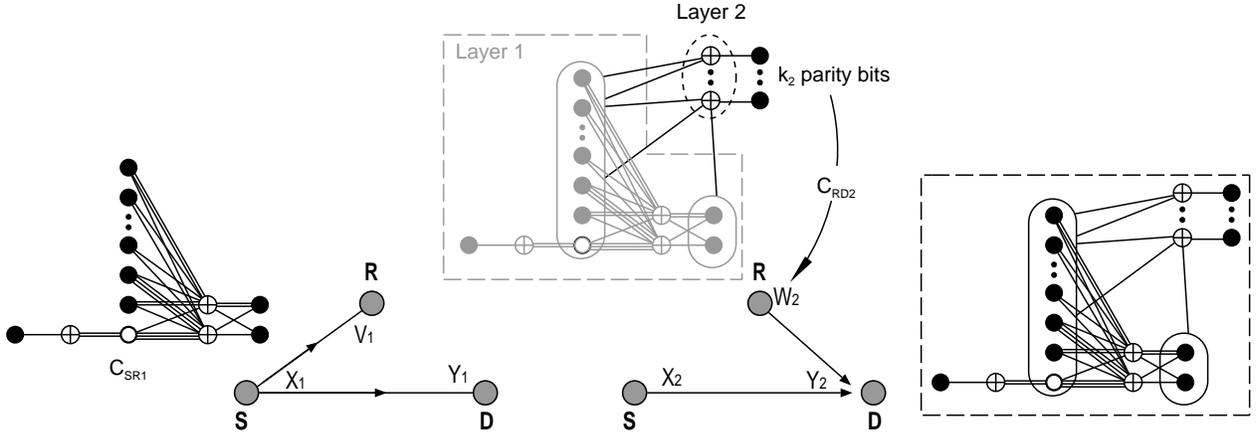} 
\caption{Bilayer expurgated protograph design for half-duplex relay with 
$R_{SD_1}=\frac{N-k_1-k_2}{N}$ and $R_{SR_1}=\frac{N-k_1}{N}$, $N$ is  
the number of variable nodes} 
\label{fig:relay_BE} 
\end{figure*}

A bilayer expurgated structure for the relay channel~\cite{Razaghi2007}
is shown in Fig.~\ref{fig:relay_BE} where a low-rate code is
constructed by adding check nodes to the graph of a high-rate
code. The bilayer graph in this case contains two layers (sets) of
check nodes and one set of variable nodes (unlike the bilayer
lengthened structure). The variable nodes together with one layer of
check nodes form the graph of a high-rate code, and the entire graph
constitutes a low-rate code. The corresponding parity check matrix has
the form
\begin{equation} 
H_{SD_1} = \begin{bmatrix} H_{SR_1}\\H_e\end{bmatrix}  \; ,
\label{eq:BE_matrix}
\end{equation} 
where $H_{SR_1}$ is the parity check matrix of a high-rate LDPC
capacity-approaching code for the source-relay link (representing the
sub-code $C_{SR_1}$ of Layer 1), and $H_e$ is the extension matrix
representing Layer 2. Layers 1 and 2 together create a
capacity-approaching code $C_{SD_1}$ for the source-destination link
as shown in Fig.~\ref{fig:relay_BE}. The source transmits a $C_{SR_1}$
codeword. The relay, after decoding the source codeword, produces $k_2$
additional ``parity'' bits using the extension matrix $H_e$,
re-encodes these $k_2$ bits with a codebook $C_{RD_2}$ and transmits to
the destination. At the destination, the $k_2$ parity bits that are
reliably detected essentially provide $k_2$ additional check values on
the source codeword from $C_{SR_1}$. This is equivalent to decoding a
$C_{SD_1}$ codeword (with a lower rate) at the SNR of the
source-destination link.

We now describe an example for the design of bilayer expurgated
codes. In the interest of brevity, we only present one construction of
bilayer codes. However, the proposed method is completely general; we
can start from any high-rate code to build a family of expurgated
codes. The starting point for the following construction is the
rate-3/4 protograph code designed in the last subsection. The
rate-$3/4$ protograph contains $13$ variable nodes (one of them
punctured) and $4$ check nodes. Low-rate bilayer codes expurgated from
this code have rates $R=\frac{13-4-n}{13-1}$, where $n$ is the number
of checks added. For each value of $n$, the new code will require a
search. As a representative sample, we concentrate on the search for
$n=1$, which yields a rate-$2/3$ code. The search space for this new
code is in the form

\begin{equation}
H^{search}_{2/3} = \left(
\begin{array}{cccccccc}
&&&&&&&\\
&&\multicolumn{4}{c}{H_{3/4}}&&\\
&&&&&&&\\\hline
0& y& x_1&x_2&  .& .& .& x_{11}
\end{array}
\right) \; ,
\label{eq:harq_x}
\end{equation}
where $y$ variable denotes the number of parallel edges connecting to
the punctured variable node and $x_i$ variables denote the number of
parallel edges connecting to remaining variable nodes except the
degree-$1$ variable node. Having a degree-$1$ variable node improves the
iterative decoding threshold of the protograph
code~\cite{Divsalar2009}.\footnote{It is well known that degree-1
nodes are not to be used in irregular LDPC codes, but their effectiveness in
structured LDPC codes is well documented~\protect\cite{Richarson_IT04}.}

The optimization process is simplified by further limiting the maximum
number of parallel edges. For example, we can limit $y\in\{1,2\}$ and $x_i\in\{0,1,2\}$. Proto-matrices of new
expurgated protographs are given by
\begin{align*} 
H^e_{2/3} = \left( 
\begin{array}{ccccccccccccc} 
\\ 
\\ 
&&&&&\multicolumn{3}{c}{H_{3/4}} 
\\ 
\\\hline 
0& 1& 0& 0& 0& 0& 0& 1& 1& 0& 0& 1& 2   
\end{array} 
\right)
\\ 
H^e_{7/12} = \left( 
\begin{array}{ccccccccccccc} 
\\ 
\\ 
&&&&&\multicolumn{3}{c}{H^e_{2/3}} 
\\ 
\\\hline 
0&1& 0& 0& 1& 0& 0& 1& 0& 0& 0& 0& 2 
\end{array} 
\right)
\\
H^e_{1/2} = \left( 
\begin{array}{ccccccccccccc} 
\\ 
\\ 
&&&&&\multicolumn{3}{c}{H^e_{7/12}} 
\\ 
\\\hline 
0&2&0&0&0&0& 0& 0& 1& 0& 1& 0& 0 
\end{array} 
\right)
\\ 
H^e_{5/12} = \left( 
\begin{array}{ccccccccccccc} 
\\ 
\\ 
&&&&&\multicolumn{3}{c}{H^e_{1/2}} 
\\ 
\\\hline 
0& 2& 0& 0& 0& 0& 0& 0& 1& 1& 0& 0& 0
\end{array} 
\right)
\\ 
H^e_{1/3} = \left( 
\begin{array}{ccccccccccccc} 
\\ 
\\ 
&&&&&\multicolumn{3}{c}{H^e_{7/12}} 
\\ 
\\\hline 
0 &2& 0& 0& 1& 0& 0& 0& 0& 0& 0& 0& 0 
\end{array} 
\right)
\end{align*} 

The iterative decoding thresholds of these codes are given in
Table~\ref{ta:proto}. The produced codes all have decoding thresholds
within $0.23$ dB of the capacity except the rate-7/12 one which has a
decoding threshold within $0.33$ dB of the capacity.

\section{Numerical results} 
\label{sec:numerical_results}

\subsection{Coding thresholds}
We design nested bilayer lengthened and expurgated protograph codes
for the half-duplex relay channel, which can operate at any two rates
with iterative decoding thresholds close to capacity, as shown in
Table~\ref{ta:proto}. For comparison, the thresholds of bilayer
lengthened and bilayer expurgated {\em irregular} LDPC codes proposed
in the literature are shown in
Table~\ref{table:ThresholdComparison}. All codes use a rate pair
($R_{SD_1}\simeq 0.5, R_{SR_1}\simeq 0.7$). Our proposed bilayer
lengthened and bilayer expurgated codes are better than the codes
proposed in~\cite{Cances2009} and within $0.1$ dB of codes reported in
other works\footnote{The rate of~\protect\cite{Razaghi2007,Cances2009}
is slightly less than 1/2, thus slightly skewing the comparison in
their favor.}. Thus, a simple design approach has yielded a family of
codes that offer a performance comparable with other highly-optimized
bilayer LDPC codes, while offering important advantages in terms of
rate-compatibility, low encoding complexity, and the fast decoding
properties of protograph codes.

\begin{table*} 
\centering 
\caption{Comparison of thresholds ($R_1\simeq 0.5,R_2 \simeq 0.7$)} 
\begin{tabular}{|c|c|c|c|c|c|c|c|c|c|} 
\hline &\multicolumn{4}{c|}{Lengthened}&
\multicolumn{5}{c|}{Expurgated}\\ \cline{2-10} &
\cite{Razaghi2007}&\cite{Cances2009} &\cite{Azmi:ISIT2008} & Our BL code
& \cite{Razaghi2007}&\cite{Azmi:ISIT2009}
&\cite{Azmi:ISIT2008}&\cite{Cances2009} & Our BE code\\ \hline Gap$_1$ &
0.164 & 0.3854 & 0.1039 &0.239 & 0.514 & 0.258 & 0.284 & 0.6323&
0.233\\ \hline Gap$_2$ & 0.120 & 0.1758& 0.0945&0.152 &0.084 & 0.084&
0.084&0.215&0.123 \\ \hline
\end{tabular} 
\label{table:ThresholdComparison} 
\end{table*} 

\subsection{Simulation Results} 

So far we have represented codes only in the form of proto-matrices
(protographs). As mentioned in Section~\ref{sec:protograph}, a
protograph code (an equivalent LDPC code) is a large derived graph
constructed by copy-and-permutation operation on protograph, a process
known as {\em lifting}.

Our protograph codes are derived from protographs in two lifting
steps. First, the protograph is lifted by a factor of $4$ using the
progressive edge growth (PEG) algorithm~\cite{Hu03_PEG} in order to
remove all parallel edges. Then, a second lifting using the
PEG algorithm was performed to determine a circulant permutation of
each edge class that would yield the desired code block
length.

In our nested bilayer lengthened codes of Section~\ref{sec:bilayerL},
the parity check matrix for the lower-rate code can be obtained by
removing certain columns from the parity check matrix of the higher rate
code, and this produces economies in the design of the decoders. In
fact, it is enough to design a decoder for the largest rate code
(9/10). To decode the lower-rate codes, the missing coded bits are
replaced by erasures at the decoder. In the same manner, the bilayer
expurgated codes generated in Section~\ref{sec:bilayerEx} only need the
decoder of the lowest rates ($1/3$). Other higher-rate codes are
decoded by ignoring redundant checks at the common decoder.

\begin{figure} 
\centering 
\includegraphics[width=3.75in]{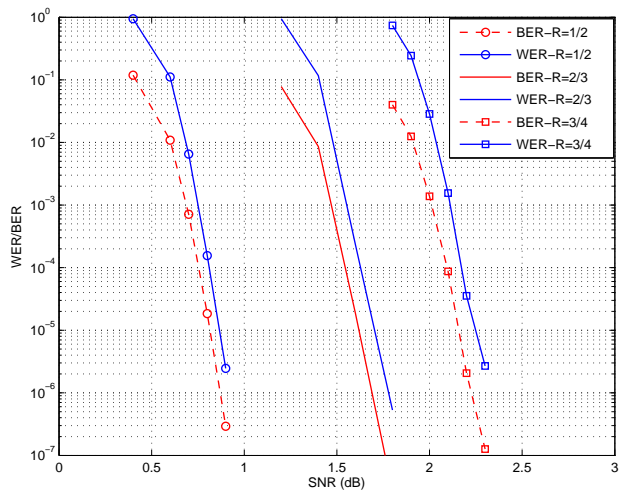} 
\caption{Performance of rate $1/2$, $2/3$, $3/4$ bilayer lengthened protograph codes designed in Section~\ref{sec:bilayerL} with
 information blocklength of 16k}
\label{fig:len_12_34}
\end{figure}

\begin{figure} 
\centering 
\includegraphics[width=3.75in]{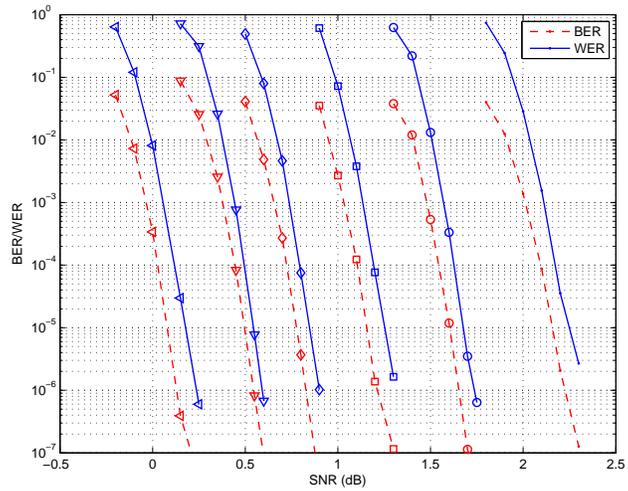} 
\caption{Performance of rate $1/3$, $5/12$, $1/2$, $7/12$, $2/3$,
  $3/4$ bilayer expurgated protograph codes (Section~\ref{sec:bilayerEx}), information block-length
 16k}
\label{fig:Ex_042_34}
\end{figure}

The performances of our bilayer lengthened protograph codes with rates
$1/2$, $2/3$ and $3/4$ over a binary-input additive white Gaussian
noise (BI-AWGN) are shown in Fig.~\ref{fig:len_12_34}. In addition,
performances of bilayer expurgated protographs with rates $1/3$, $5/12$,
$1/2$, $2/3$ and $3/4$ are shown in Fig.~\ref{fig:Ex_042_34}. All
codes are simulated with the information block-length of $16$k. To
construct codes with this $16$k information block-length, the
protographs for the lengthened codes are first lifted by a factor of
$4$, and subsequently again by factors $1365$, $683$, and $455$, for
codes of rate $1/2$, $2/3$, and $3/4$, respectively. A similar two-step
lifting is used for the expurgated codes. 

The decoder is a standard message passing decoder where the maximum
number of iterations is set to $200$. Log-likelihood ratio (LLR) clipping and other decoding
parameters are according to~\cite{Hamkins:IPN2011}. No error floors
were observed down to the word error rate (WER) $10^{-6}$. The gap to capacity of these
codes is within $0.6$ dB of their iterative decoding threshold and
within $0.8$ dB of their capacity limits at $WER\approx 3 \times 10^{-6}$.

We now consider a half-duplex relay channel with information
block-length of $16380$ bits.  We assume the SR link supports rate
$R_{SR_1}=3/4$ and the SD link supports rate $R_{SD_1}=1/2$. For this
example, the time division $t=0.75$ is chosen for the source and
relay.
The rate needed in the RD link is
$R_{RD_2}=3/4$~\cite[Eq.  5]{Chakrabarti2007}. This is the same as the
rate of SR code but with a different codeword-length. Under these
conditions, the achievable rate of the relay channel is
$0.5625$~\cite[Eq. 3]{Chakrabarti2007}.

\begin{figure} 
\centering 
\includegraphics[width=3.75in]{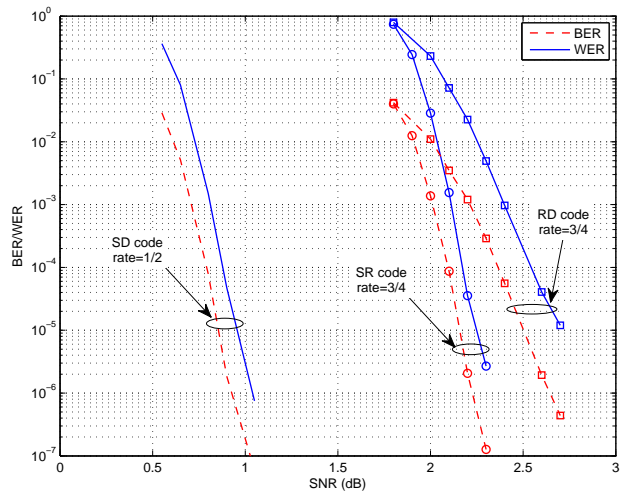} 
\caption{Performance of component codes for a relay channel:
 $R_{SD_1}$=1/2, $R_{SR_1}$=3/4 and $R_{RD_2}=3/4$ using bilayer expurgated structure 
 (Section~\ref{sec:bilayerEx}) with $\alpha=1.4$ dB and $\beta=1.6$ dB}
\label{fig:Ex_12_34}
\end{figure}

We now consider the bilayer expurgated coding scheme
(Fig.~\ref{fig:relay_BE}) with information blocklength
$16380$. $C_{SR_1}$ is implemented by the rate-$3/4$ protograph given
in Section~\ref{sec:bilayerL}.  The code $C_{RD_2}$, which protects
additional parities generated at the relay, has information
blocklength $5460$. $C_{RD_2}$ is constructed from the same protograph
as $C_{SR_1}$, but with a shorter length. The destination decodes a
$C_{SR_1}$ codeword plus these additional parity values, which
constitutes a codeword of $C_{SD_1}$ in the same manner
as~\cite{Razaghi2007}.  The word error rate (WER) and bit error rate (BER) performances of these codes
are shown in Fig.~\ref{fig:Ex_12_34}. The shorter block length of the
rate-$3/4$ RD code has resulted in a wider waterfall region,
highlighting a weakness of the successive decoding approach at smaller
block lengths.

\begin{figure} 
\centering \includegraphics[width=3.75in]{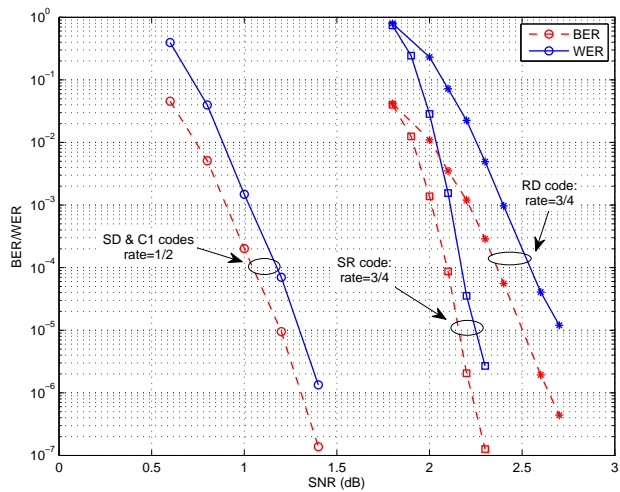}
\caption{Performance of component codes for a relay channel:
 $R_{SD_1}$=1/2, $R_{SR_1}$=3/4, $R_{C_1}=1/2$ and $R_{RD_2}=3/4$ using bilayer lengthened
 structure in Section~\ref{sec:bilayerL} with $\alpha=1.4$ dB and
 $\beta=1.6$ dB}
\label{fig:BL_12_34}
\end{figure}

The bilayer lengthened structure involves four codes
(Fig.~\ref{fig:relay_BL}). Using the above relay channel parameters,
$C_{SR_1}$ has the information block-length $16380$ and codeword block
length $21840$. Each codeword of $C_{SR_1}$ is composed of a
$C_{SD_1}$ codeword with blocklength $10920$ and an extension of
$10920$ bits derived via the extension matrix $H_e$ from
Eq.~\eqref{eq:BL_matrix}. The relay, after decoding and recovering the
codeword, multiplies the extension bits by the parity check matrix of
a rate-$1/2$ code $C_1$ to calculate a syndrome of length
$5460$. $C_1$ is lifted from the rate-$1/2$ protograph given in
Eq.~\eqref{eq:B12_new}. The syndrome is then encoded by $C_{RD_2}$,
transmitted and decoded at the destination to recover the syndrome.
$C_{SR_1}$ and $C_{RD_2}$ are similar to their counterparts in the
bilayer expurgated scheme.  Also, $C_{SD_1}$ and $C_1$ have the same
code structure and blocklength.  The WER and BER performances are
shown in Fig.~\ref{fig:BL_12_34}. Performance of $C_{SD_1}$ in the
bilayer-lengthening structure is worse than in the bilayer
expurgated structure due to its shorter blocklength.

Although the example above was coached in terms of orthogonal
transmissions, it also applies to a non-orthogonal system with
correspondingly lower source and relay powers (see
footnote~\ref{footnote:2}).

\subsection{End-to-End Performance of Relay Coding Systems} 
\label{sec:system_evaluation}


Many works in contemporary literature represent the performance of
relay channel codes by illustrating two component codes: the
source-relay code, and the source-destination code. In this approach
it is implicitly assumed the relay-destination code is an ideal
code. However, the relay-destination code often operates at a smaller
block length, thus it has a wider waterfall and may very well be the
bottle neck for the entire system, therefore providing the thresholds
or even simulations for only the two other sub-codes may not always be
fully illuminating of the overall performance. We believe there is
need for a comprehensive end-to-end performance metric.


In general the end-to-end relay channel error is a function of the
three SNRs of its constituent channels. But showing this dependence
requires a four-dimensional plot, which is not practical. We construct
a simple but useful error plot by assuming
\begin{equation}
SNR_{SR}=SNR_{SD}+\alpha \qquad SNR_{RD}=SNR_{SD}+\beta \; ,
\label{eq:SNR-relation}
\end{equation}
where $\alpha$ and $\beta$ are fixed constants.  The affine
relationship of the SNRs allows them to be displayed on the same axis,
producing a simple end-to-end error plot which can be thought of as a
2-dimensional slice of the general four-dimensional error
characterization of the relay channel. $\alpha$ and $\beta$ can be
carefully chosen (as shown below) to generate a reasonable and
insightful overall representation of the performance of the relay
channel that does not ignore or trivialize any of the parameters of
the system.

To simplify the simulations, we develop a tight bound on the end-to-end
error based on component errors in the system. For the purposes of
exposition we concentrate on the expurgated-bilayer codes
(Figure~\ref{fig:relay_BE}), where three types of errors may happen in
the system: $E_R$ is the error event at the relay, $E_{RD}$ is the error
in the decoding of the ``extra parity'' arriving from the relay to the
destination, and finally, $E_D$ is the error event at the destination in
the final decoding of the source message, and the complement of event $E$ is shown with $\overline{E}$. The bound  is shown in Eq.~\eqref{eq:relay_bound}. 
\begin{align} 
P_e &= P(E_D|\overline{E_R})P(\overline{E_R}) + P(E_D|E_R)P(E_R) \\ 
&=
\big[P(E_D|\overline{E_{RD}},\overline{E_R})P(\overline{E_{RD}})+
 P(E_D|E_{RD},\overline{E_R})P(E_{RD})\big]P(\overline{E_R}) +
P(E_D|E_R)P(E_R) \\&\leq P(E_D|\overline{E_{RD}},\overline{E_R}) +
P(E_{RD}) +P(E_R)
\label{eq:relay_bound} 
\end{align}   
It is easy to see the bound is tight because component codes are used
in a regime where their errors are, conservatively, no more than
$10^{-2}\sim 10^{-3}$, therefore with a very good approximation
$P(\overline{E_R})\approx 1$, $P(\overline{E_{RD}})\approx 1$,
$P(E_D|E_R)\approx 1$ and $P(E_D|E_{RD},\overline{E}_R)\approx 1$. 

\begin{figure}
\centering \includegraphics[width=3.75in]{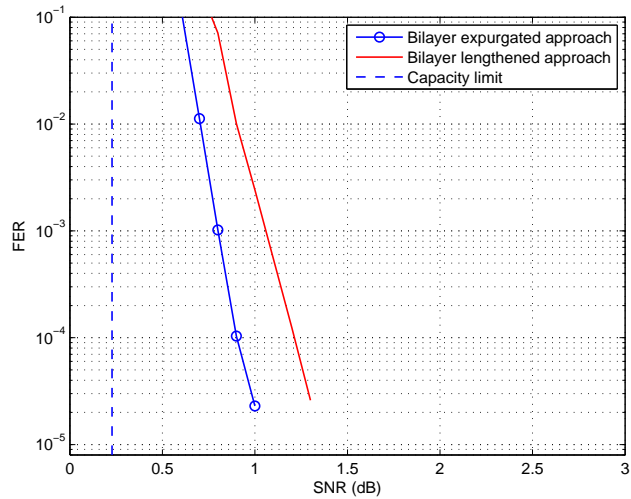} 
\caption{End-to-end performance bounds for relay coding schemes in
 Fig.~\ref{fig:BL_12_34} and Fig.~\ref{fig:Ex_12_34} with $\alpha=1.4$
 dB and $\beta=1.6$ dB }
\label{fig:bound_BL_BE}
\end{figure} 

We are now ready to calculate the values of $\alpha$ and $\beta$,
which will determine how the individual code error characteristics are
combined to produce the end-to-end performance. We are interested in
cases where none of the component errors dominate the others, because
when one of the link errors dominates, the performance will be
essentially determined by the code on the dominating link, which is
already known from the literature on point-to-point channels. So the
interesting case is when no link errors dominate, i.e. the waterfall
regions of the component curves coincide at their starting points
(approximately the threshold).

The performance of two bilayer codes with a rate pair
($R_{SD}=1/2,R_{SR}=3/4$), $\alpha=1.4$ dB, $\beta=1.6$ dB and time
sharing $t=0.75$ are shown in Fig.~\ref{fig:bound_BL_BE}. To
understand how far is the end-to-end performance of our codes from
theoretical limits, we follow the convention of point-to-point
channels and produce a value of SNR at which a random code of infinite
length and rate similar to the code under study is capable of
theoretically supporting error-free communication. Similarly to the
point-to-point case, this involves inverting the capacity formula by
inserting the rate and extracting the corresponding SNR. To produce a
single SNR value, we use Eq.~\eqref{eq:SNR-relation}, assume that each
of the component codes is at the rate that the mutual information of the link
supports, and use the relay mutual information formulas
in~\cite{Chakrabarti2007}. Using these parameters, the relay channel
of the above example has a throughput of $0.5625$ and the code rates
correspond to the limiting SNR of $SNR_{SD}=0.225$ dB. As seen in this
figure, the gap-to-capacity of bilayer expurgated and bilayer
lengthened relay coding schemes are about $0.7$ dB and $1.2$ dB
respectively at the frame error rate (FER) of $2 \times 10^{-5}$. The end-to-end error bound of the bilayer
lengthened code is worse than that of the bilayer expurgated code
because of the block-length issues. This phenomenon has also been reported by
several previous works including~\cite{Razaghi2007,Cances2009}.

\section{Desing of Bilayer Codes for Two-Relay Channels} 
\label{sec:two_relay} 

\begin{figure*} 
\centering 
\includegraphics[width=6.5in]{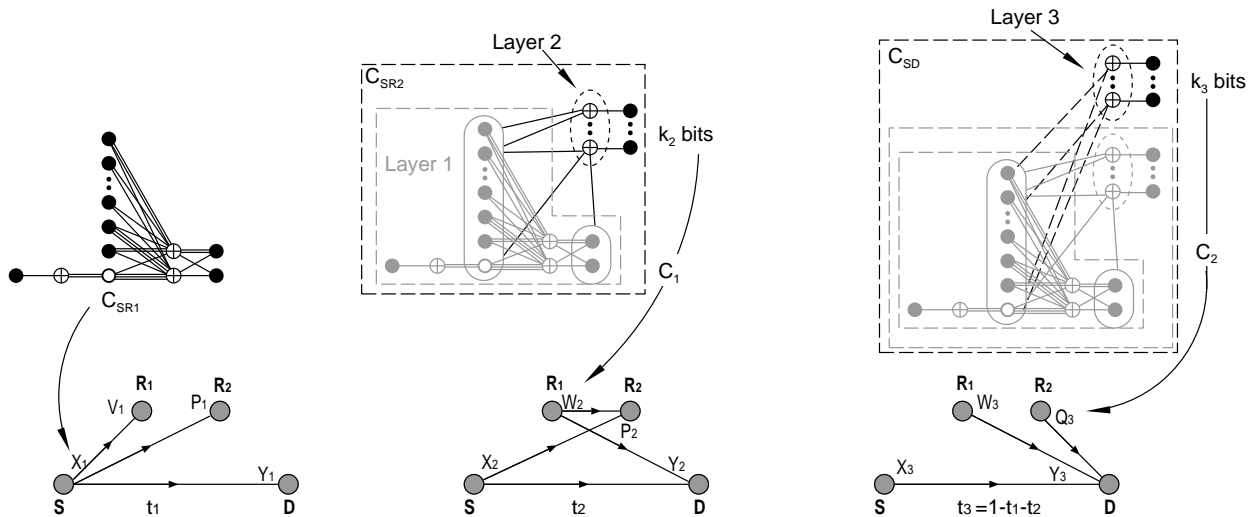} 
\caption{The expurgated coding structure for the half-duplex two-relay channel} 
\label{fig:two_relay} 
\end{figure*} 

In a practical two-relay scenario (Fig.~\ref{fig:two_relay}) it is
likely that the relays do not have precisely and deterministically
identical channels to the source, destination, and each
other. Therefore one relay is likely to be ``stronger'' and decode
first, then this relay will be able to help another relay to decode,
and then the two relays together will assist the destination.

We consider transmissions in three time slots for the source and two
relays. The transmitted signals from the source (S) and two relays
($R_1$ and $R_2$) are denoted with $X$, $W$ and $Q$, respectively, and
the received signals at the destination and two relays are denoted
with $Y_i$, $V_i$ and $P_i$, respectively, where $i$ indicates
the time slot index. In the first time slot, the source encodes its
message using the code $C_{SR_1}$. The first relay decodes the source
message, but the second relay and the destination cannot (yet). During
the second time slot, the first relay generates $k_2$ parity bits
using the sub-graph denoted Layer~2,\footnote{Like the one-relay case,
these bits are the syndrome of a parity check extension matrix.} then
encodes these parity bits using another LDPC code $C_1$. The second
relay, after decoding the $k_2$ parity bits, decodes the source
message. Then, the second relay computes $k_3$ parity bits using the
subgraph denoted Layer~3, encodes the $k_3$ bits with another LDPC
code $C_2$, and transmits to destination. The destination decodes
$C_1$, $C_2$, and finally the source message with the help of
$k_2+k_3$ additional parity bits from two relays. The achievable rate
using the above strategy is a special case of the achievable rate
in~\cite{Xie:IT05}.

For demonstration, consider an example where $R_1$, $R_2$ and $D$ can
reliably receive source signals with a rate $R_{SR_1}=3/4$,
$R_{SR_2}=7/12$ and $R_{SD}=1/3$, respectively. The $R_1$-to-$D$
channel supports $R_{R_1D}= 1/2$. Assuming $t_1=0.6$, $t_2=0.2$ and
$t_3=0.2$, the achievable rate of the two-relay channel is
$0.45$~\cite{Chakrabarti2007}. The rate of $C_1$ and $C_2$ are $1/2$
and $3/4$, respectively. Information block length is $16380$.
$C_{SR_1}$, $C_{SR_2}$ and $C_{SD}$ are constructed from protographs
given in Section~\ref{sec:bilayerEx}. $C_{SR2}$ is rate-$7/12$ and
$C_{SD}$ is rate-$1/3$, resulting in $k_2=3640$ and
$k_3=5460$. Although the explanation of this example was in terms of
orthogonal transmissions, the precise same example also applies to a
non-orthogonal (beamforming) relay channel with correspondingly lower
source and relay power. 

\begin{figure} 
\centering 
\includegraphics[width=3.75in]{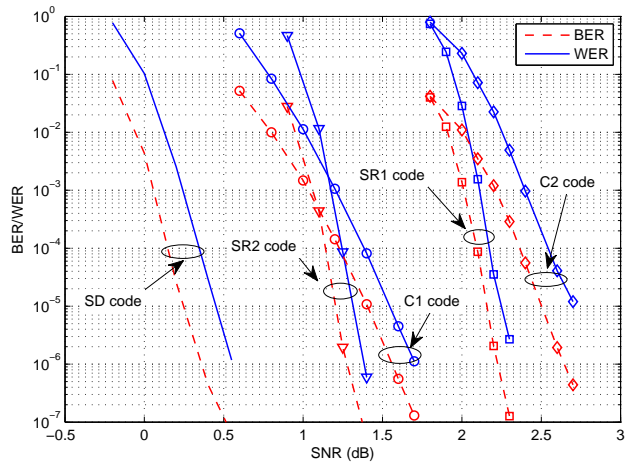} 
\caption{Performances of component codes used in a two-relay channel:
 $R_{SR_1}=3/4$, $R_{SR_2}=7/12$, $R_{SD}=1/3$, $R_{C_1}=1/2$ and
 $R_{C_2}=3/4$. $C_{SR_1}$, $C_{SR_2}$ and $C_{SD}$ all have codeword
 blocklengths of $23660$, while $C_1$ and $C_2$ have codeword
 blocklengths of $7280$.}
\label{fig:two_relay_code}
\end{figure}

$C_1$ arises from the protograph in Eq.~\eqref{eq:B12_new} and $C_2$
is the same code as $C_{SR_1}$, but with a shorter blocklength.
Fig.~\ref{fig:two_relay_code} shows the performance of $C_{SR_1}$,
$C_{SR_2}$, $C_{SD}$, $C_1$ and $C_2$, which operate within $0.6$ dB,
$0.7$ dB, $0.9$ dB, $1.5$ dB and $1$ dB, respectively of their capacity
limits at $WER = 10^{-5}$. As expected, $C_1$ is the bottleneck in
this two-relay channel because of its short blocklength. We note that
this gap is an outcome of a relatively short overall block length and
a successive decoding approach, both of which are practical
considerations and can be relaxed.

\begin{remark}
In the multi-relay scenario, there are multiple options for the
decoding and transmission at relays. For example, two relays can
cooperatively beamform towards a destination. The required codes in
this case are fundamentally similar to the one-relay channel,
therefore we do not consider them separately in this paper.
\end{remark}

\begin{remark}
In the multi-relay scenario the transmission time is divided multiple
times which shortens the component block-lengths. Therefore we
did not pursue a multi-relay generalization of the bilayer {\em
lengthened} structure due to its susceptibility to short block length
effects.
\end{remark}

\begin{remark}
Optimization of the time division to maximize the achievable rates has
been pursued in papers on the capacity of relay channel~\cite{Madsen:vtc02} and
can be extended to the multi-relay scenario. Depending on the channel
gains, the optimal length for one of the time slots may be zero, in
which case one of the relays must be shut off.
\end{remark}

\begin{remark}
The developments throughout this paper apply to either orthogonal relaying or
to non-orthogonal relaying where concurrent transmissions use the same
codebook. 
\end{remark}

\section{Discussion and Conclusion} 
\label{sec:conclusion} 

This paper presents a simple approach for constructing relay coding
schemes based on bilayer lengthened and bilayer expurgated protograph
codes which perform within a fraction of dB of the capacity. The
proposed codes allow easy design, flexibility in matching to
various relay channel conditions and low encoding complexity and
can be extended to multi-relay networks. A framework for
end-to-end performance evaluation of the relay codes is also provided.

Nested protograph codes have also been used for the
point-to-point channel~\cite{Thuy:Tcom10} where, like the present
work, multi-component protograph codes have been considered. However, in relay
channels the component codes are emitted by different transmitters,
unlike the point-to-point case, and see different channels with
different SNRs. Thus, relay networks represent a different scenario
whose challenges are addressed by a bilayer
structure~\cite{Razaghi2007} for the overall code and a coset code at
the relay, techniques that have no counterpart in the point-to-point
case.

\section*{Acknowledgment}
This research was supported in part by the VOSP program from the
Ministry of Education and Training, Vietnam, and in part by the grant
009741-0084-2007 from THECB. This research was in part carried out at
the Jet Propulsion Laboratory, California Institute of Technology,
under a contract with NASA.

\bibliographystyle{IEEEtran} 
\bibliography{IEEEabrv,relay_ref}

\end{document}